# Superconductivity of CH$_4$-intercalated H$_3$S under high pressure

Mingyang Du[1], Zihan Zhang[1], Tian Cui[2,1,*], Defang Duan[1,*]

[1] College of Physics, Jilin University, Changchun 130012, People's Republic of China

[2] Institute of High Pressure Physics, School of Physical Science and Technology, Ningbo University, Ningbo, 315211, People's Republic of China

KEYWORDS: Hydrides, High pressure, Superconductivity, first principles calculation.

**ABSTRACT**

The discovery of the high temperature superconducting state in compounds of hydrogen, carbon and sulfur with the critical temperature ($T_c$) of 288 K at high pressure is an important milestone towards room-temperature superconductors. Here, we have extensively investigated the high-pressure phases of $CS_2H_{10}$, and found four phases $Cmc2_1$, $P3m1$, $P\text{-}3m1$ and $Pm$. Among them, $P3m1$ can be dynamically stable at pressure as low as 50 GPa, and $Cmc2_1$ has high $T_c$ of 155 K at 150 GPa. Both of $Cmc2_1$ and $P3m1$ are host-guest hydrides, in which $CH_4$ molecules are inserted into $Im\text{-}3m\text{-}H_3S$ and $R3m\text{-}H_3S$ sublattices, respectively. Their $T_c$ is dominated by the $H_3S$ lattice inside. The insertion of $CH_4$ greatly reduces the pressure required for the stability of the original $H_3S$ lattice, but it has a negative impact on superconductivity that cannot be ignored. By studying the effect of $CH_4$ insertion in the $H_3S$ lattice, we can design hydrides with $T_c$ close to that of $H_3S$ and a greatly reduced pressure required for stability.

**Introduction**

In the Ginzburg's list of "especially important and interesting problems"[1], room temperature superconductivity and metallic hydrogen ranked second and third, respectively. Both problems have attracted solid physicists for decades. In recent years, these problems seem to have a common solution: high-pressure hydrides. According to BCS theory[2], the high superconducting transition temperature ($T_c$) requires higher electronic density of states and phonon frequencies near the Fermi level. Metallic hydrogen formed under high pressure is expected to be a high temperature superconductor because of large vibration frequency[3, 4], but theoretical and experimental results show that metallic hydrogen requires extremely high pressure[5, 6]. Hydrogen-rich compounds can be an alternative way to find high-temperature superconductors, because the "chemical pre-compression" greatly reduces the metallization pressure[7, 8].

Theoretical investigations have predicted that $H_3S$ is a high temperature superconductor with $T_c$ of about 200 K at high pressure [9, 10], which has been confirmed by experiments, with the $T_c$ of 203 K[11, 12]. This observation indicates the feasibility of finding high $T_c$ in hydrogen-rich materials, attracting people to quest more hydrogen-rich superconductors under high pressure. Metal hydrides have been the focus of research [13-21]. Some of them have been experimentally confirmed, such as $LaH_{10}$[22, 23], $YH_6$[24], $YH_9$[25], $CaH_6$[26] and $ThH_{10}$[27]. Recently, Snider et al. observed the high-$T_c$ state with 288 K in compounds of hydrogen, carbon and sulfur at pressure of 267 GPa, which is a milestone towards room-temperature superconductors[28]. Unfortunately, its stoichiometry and crystal structure are still uncertain. In fact, not only this room temperature superconducting phase, the entire structural evolution of C-S-H system under high pressure are lack, which makes subsequent research difficult. Therefore, investigation of the stoichiometry and structural evolution of C-S-H system is necessary to provide useful complementary information for experimental observation.

When Li et al. studied $Pm$-3$m$-H$_3$SXe with a similar structure to $Im$-3$m$-H$_3$S, they found that the "H$_3$S" host lattice is a key factor affecting the $T_c$[29]. Therefore, the room temperature superconducting C-S-H compound was considered to be C-doped H$_3$S for a period of time. Later, Ge et al. did find that 3.8% C (H$_3$S$_{0.962}$C$_{0.038}$) can raise the $T_c$ of H$_3$S to 289 K at 260 GPa[30]. However, the recent two experiments on the high-pressure structure and composition of the C-S-H system reported that its structure is complex and different from the common $Im$-3$m$-H$_3$S[31, 32]. According to the experimental data, the C-S-H compound should be much richer in hydrogen than H$_3$S. A metastable hydrogen-rich host-guest CSH$_7$ compound with similar structures $I$-43$m$, $Cm$, $R$3$m$, and $Pnma$ were found before, where C and H form methane molecules CH$_4$ which was inserted into the H$_3$S sublattice[33]. The electron-phonon coupling calculations reveal superconductivity with $T_c$'s as high as 194 K at 150 GPa[34]. Most recently, Wang et al. studied the C-S-H system more extensively[35] and believe that the structure containing more carbon and hydrogen is unlikely to be a low-enthalpy phase and easily decomposes into H$_2$, H$_3$S, or CH$_x$. In the ternary structures they explored, only CS$_2$H$_{10}$ has a lower enthalpy (-65 meV/atom).

Here, we have extensively investigated the high-pressure phases of CS$_2$H$_{10}$ and found $Cmc2_1$, $P$3$m$1, $P$-3$m$1 and $Pm$ phases. The $Cmc2_1$ is a host-guest structure in which CH$_4$ molecules are inserted into the H$_3$S sublattice, which is a potential high-temperature superconductor with $T_c$ of 155 K at 150 GPa. The H$_3$S lattice makes a decisive contribution to superconductivity, and the CH$_4$ molecules in the middle of the H$_3$S sublattice can greatly reduce the pressure required for the stability of the H$_3$S lattice. This discovery can not only help us further understand the evolution process of the C-S-H system within 50-200 GPa, but also provide useful guidance on how to reduce the stable pressure of high-temperature superconducting hydrides.

**Computational details**

Ab initio random structure searching (AIRSS) technique[36, 37] and ab initio calculation of the Cambridge Serial Total Energy Package (CASTEP)[38] were used to predict the candidate crystal structures of $CS_2H_{10}$. The plane-wave cut-off energy of 300 eV and the Brillouin zone sampling grid spacing of $2\pi \times 0.07$ Å$^{-1}$ were selected. The generalized gradient approximation (GGA) with the Perdew-Burke-Ernzerhof (PBE) parametrization[39] for the exchange-correlation functional and on-the-fly (OTF) generation of ultra-soft potentials were used for the structure searching.

The Vienna ab initio simulation program (VASP)[40] was used for structural relaxation and calculations of enthalpies and electronic properties. The projector augmented plane-wave (PAW) potentials[41] with an energy cutoff of 1000 eV and Monkhorst-Pack (MP)[42] meshes for Brillouin zone sampling with resolutions of $2\pi \times 0.03$ Å$^{-1}$ were used to ensure that all enthalpy calculations are well converged to less than 1 meV per atom.

The Quantum-ESPRESSO[43] was used in phonon and electron−phonon calculations. Ultra-soft potentials were used with a kinetic energy cut-off of 90 Ry. The k-points and q-points meshes in the first Brillouin zone are 9×9×9 and 3×3×3 for *Cmc2₁*, 9×9×15 and 3×3×5 for *P3m1*, 9×9×15 and 3×3×5 for *P-3m1*, 6×8×12 and 2×4×4 for *Pm*. The superconducting transition temperatures of these structures are estimated through the Allen−Dynes-modified McMillan equation (A-D-M) with correction factors[44, 45] and self-consistent solution of the Eliashberg equation (scE)[46].

**Results and discussion**

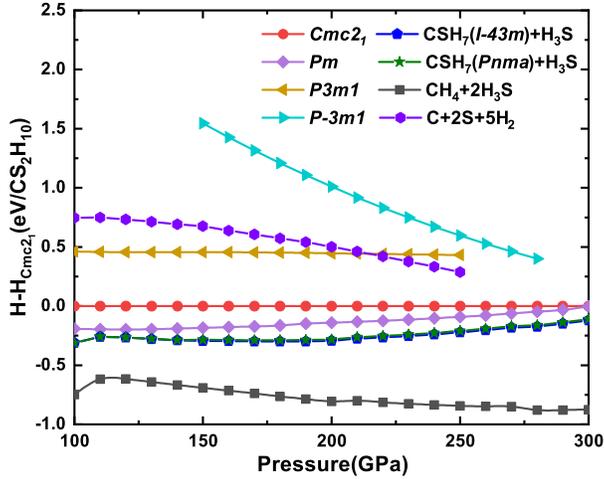

Fig. 1. Calculated enthalpies per $CS_2H_{10}$ unit as the function of pressure for our predicted candidate structures, $CSH_7$ ($I$-$43m$[33] and $Pnma$[34]) + $H_3S$ ($Cccm$, $R3m$ and $Im$-$3m$[9]), molecular assemblage $CH_4$ ($Cmcm$ and $Pnma$[47]) + $2H_3S$, and the assemblage C ($Fd$-$3m$[48]) + 2S ($β$ -Po[49]) + $5H_2$ ($P6_3/mc$, $C2/c$ and $Cmca$-12[50]) relative to the $Cmc2_1$ phase.

We performed detailed structure searches focusing on $CS_2H_{10}$ with 1 to 4 formula units at pressure of 100-300 GPa. Four new structures are uncovered: $Cmc2_1$, $P3m1$, $P$-$3m1$ and $Pm$, and their enthalpy curves are shown in Fig. 1. The $Pm$ and $Cmc2_1$ phase emerged as being stable relative to the elements in the pressure range considered, but their enthalpy is higher than $CH_4$ + $2H_3S$, and slightly higher than $CSH_7$ + $H_3S$. The calculation of the phonon spectrum shows that $P3m1$ can be dynamically stable at a pressure as low as 50 GPa. When the pressure exceeds 220 GPa, its energy is higher than the elemental substance and begins to decompose. The enthalpy of $P$-$3m1$ is much higher than other phases at 150 GPa, but its enthalpy decreases rapidly as the pressure increases, it may be more stable than other structures at pressures above 300 GPa. Then, we calculated the phonon spectrum of these structures at different pressures (see Fig. S1). As a result, there are no imaginary frequency modes in the entire Brillouin zone, which proves the dynamic stability of all four structures. It is expected that these phases could be realized experimentally.

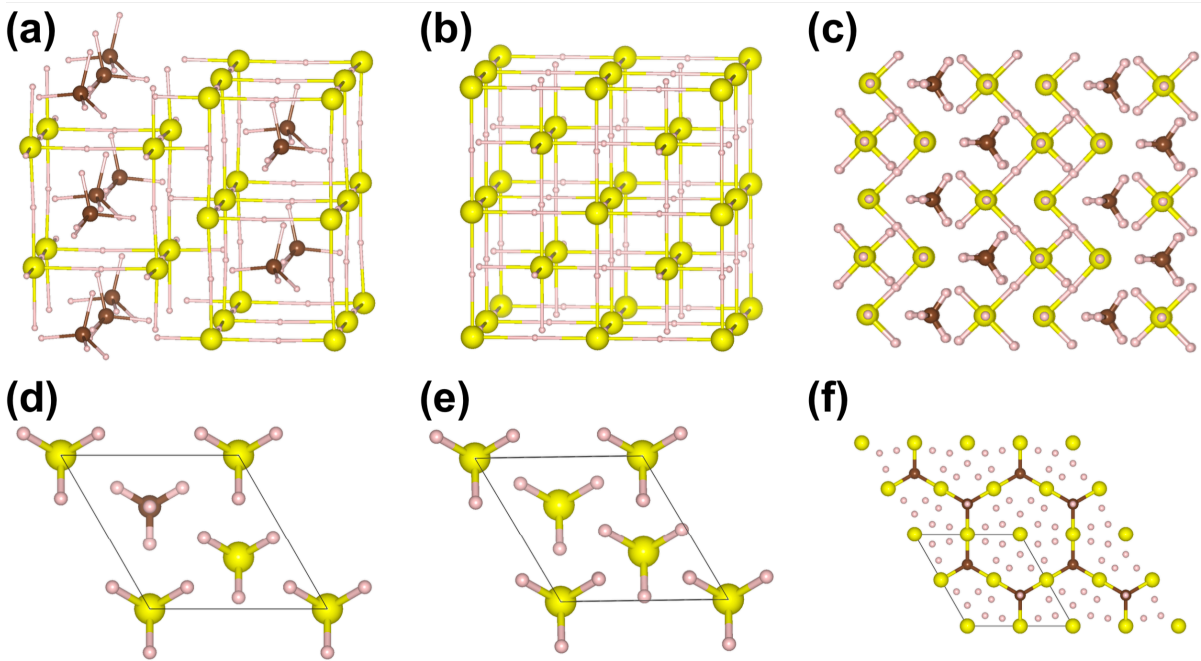

Fig. 2. (a) $Cmc2_1$-$CS_2H_{10}$ at 150 GPa, (b) $Im$-$3m$-$H_3S$ at 200 GPa, (c) $Pm$-$CS_2H_{10}$ at 200 GPa, (d) $P3m1$-$CS_2H_{10}$ at 50 GPa, (e) $R3m$-$H_3S$ at 130 GPa and (f) $P$-$3m1$-$CS_2H_{10}$ at 100 GPa. The brown, yellow, and pink spheres denote C, S, and H atoms, respectively.

The predicted crystal structures of $CS_2H_{10}$ are shown in Fig. 2, and their structure parameters are listed in Table S1 of the supplementary information. The $Im$-$3m$-$H_3S$ is characterized by S atoms located at a simple body-centered cubic lattice (bcc) and H atoms located symmetrically between S atoms. From Fig. 2a and b, it can be seen that the $H_3S$ in $Cmc2_1$ has the same lattice as $Im$-$3m$-$H_3S$, with slight distortion. The regular octahedron $SH_6$ in $Im$-$3m$-$H_3S$ was replaced with $CH_4$, which is similar to the previously reported $CSH_7$. The structure of $Pm$ is also formed by inserting $CH_4$ molecules into the $H_3S$ lattice, except that the replacement position of $CH_4$ is different from $Cmc2_1$ (see Fig. 2c). The $P3m1$ phase is composed of $CH_4$ molecules and $SH_3$ molecules, and the arrangement of $SH_3$ molecules is the same as that of $R3m$-$H_3S$ (see Fig. 2d and e). The structure of *P*-3*m*1 is composed of CS$_2$ pleated hexagonal framework and hydrogen atoms (see Fig. 2f).

Table 1. The calculated EPC parameter (λ), logarithmic average phonon frequency ($\omega_{log}$), superconducting critical temperature f$_1$f$_2$T$_c$ using Allen-Dynes modified McMillan equation and T$_c^{scE}$ using the Self-consistent solution of the Eliashberg equation.

| Phase | Pressure (GPa) | λ | $\omega_{log}$ (K) | f$_1$f$_2$T$_c$ (K) | T$_c^{scE}$ (K) |
|---|---|---|---|---|---|
| *Cmc*2$_1$ | 150 | 2.01 | 799 | 128-141 | 142-155 |
| *P*3*m*1 | 50 | 1.51 | 651 | 77-86 | 85-95 |
| *P*3*m*1 | 100 | 1.21 | 916 | 79-90 | 84-95 |
| *P*-3*m*1 | 100 | 0.63 | 1113 | 22-31 | 22-30 |
| *Pm* | 200 | 0.69 | 1386 | 36-47 | 39-50 |

the Coulomb pseudopotential µ* = 0.10 and 0.13.

To examine the superconductivity in predicted four structures of CS$_2$H$_{10}$, we calculate the electron-phonon coupling constant (λ) and average phonon frequency ($\omega_{log}$) as shown in Table S5. For *Cmc*2$_1$, λ can reach 2.01 indicating that it is a strong electron-phonon coupling superconductor. But the λ of *Cmc*2$_1$-CS$_2$H$_{10}$ is lower than that of *I*-43*m*-CSH$_7$ (λ = 3.64 at 100 GPa), *Pnma*-CSH$_7$ (λ = 3.06 at 150 GPa), *R*3*m*-CSH$_7$ (λ = 2.47 at 150 GPa[34]) and *Im*-3*m*-H$_3$S (λ = 2.19 at 200 GPa), which is not beneficial for superconductivity. The logarithmic average phonon frequency ($\omega_{log}$= 799 K) of *Cmc*2$_1$-CS$_2$H$_{10}$ is higher than that of *I*-43*m*-CSH$_7$ ($\omega_{log}$ = 395 K) and *Pnma*-CSH$_7$ ($\omega_{log}$ = 672 K). Furthermore, we estimated the *T$_c$* of *Cmc*2$_1$ to be 128 - 141 K at 150 GPa using Allen-Dynes modified McMillan equation[45] with typical values of the Coulomb pseudopotential µ* = 0.13 - 0.1. The calculation of the phonon spectrum shows that *P*3*m*1 can be dynamically stable at a pressure as low as 50 GPa, with *T$_c$* of 85-95 K. The *T$_c$* of *P*-3*m*1 and *Pm* is relatively low (< 100 K), as shown in Table 1. To better describe the systems with strong electron-phonon coupling (i.e., for λ > 1.5), we evaluate the *T$_c$* of these structures by the self-consistent solution of the Eliashberg equation[46]. The *T$_c$* of *Cmc*2$_1$ can reach 142-155 K at 150 GPa, which is close to the $T_c$ of C-S-H compound at 150 GPa measured by Snider et al. in the experiment[28]. However, the gap between $CS_2H_{10}$ and C-S-H compounds rapidly increases in the higher pressure range. Except for $Cmc2_1$ with a larger λ, the $T_c$ of other phases did not increase significantly.

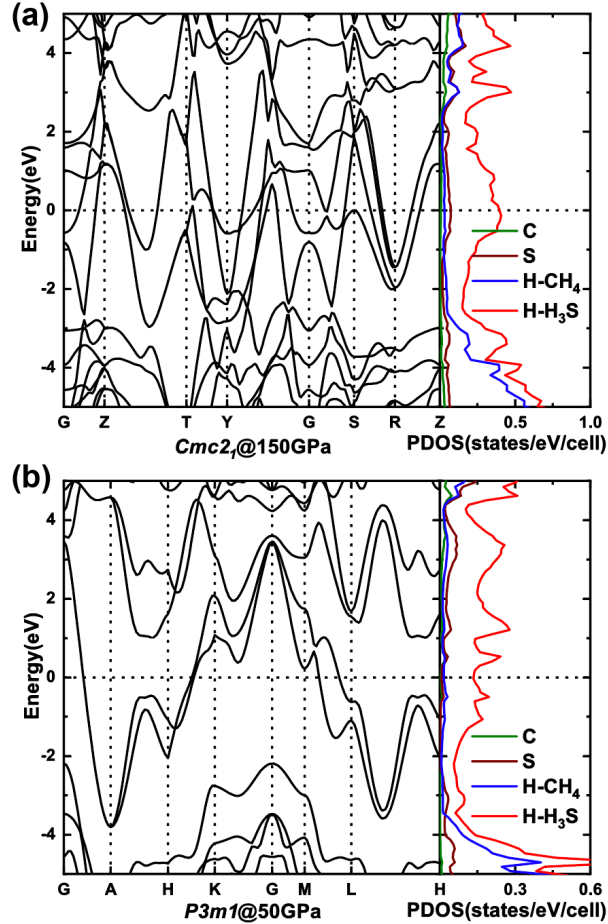

Fig. 3. Electronic band structure and projected density of states (PDOS) of (a) $Cmc2_1$ at 150 GPa and (b) $P3m1$ at 50 GPa.

In order to further explore the origin of superconductivity in $Cmc2_1$ and $P3m1$, we calculated their electronic properties. The electronic band structure shows that they are all metallic phases (see Fig. 3). It can be seen that the decisive contribution near the Fermi surface is the H in the H$_3$S lattice, and the H in CH$_4$ contribute only to the deep energy levels (< -3 eV). Unlike H$_3$S, S in CS$_2$H$_{10}$ contributes little to the Fermi surface, which is unfavorable for the superconductivity of covalent hydrides. Therefore, in these two types of CS$_2$H$_{10}$, the superconductivity is driven by H on H$_3$S, and the insertion of CH$_4$ affects the strong hybridization of the orbitals of S and H atoms, resulting in a decrease in $T_c$.

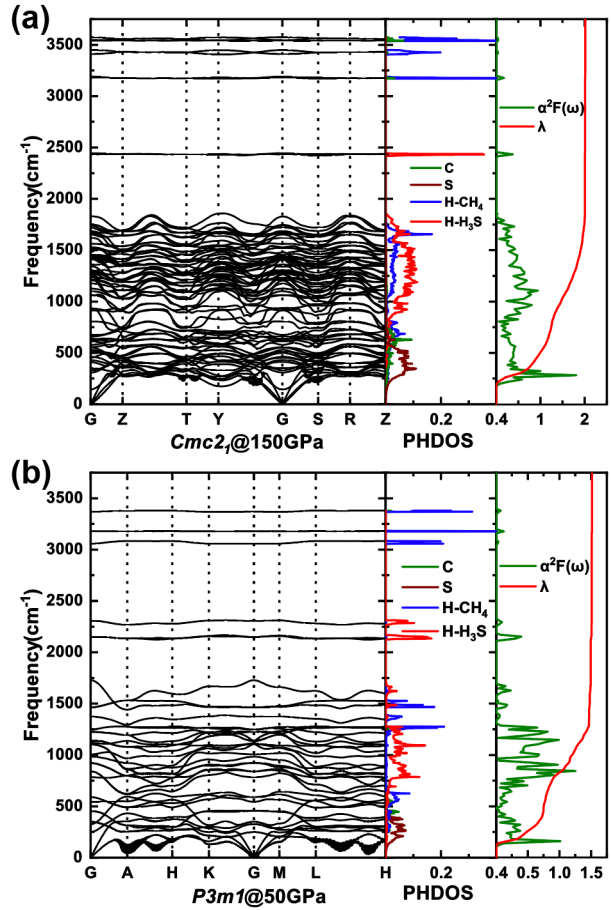

Fig. 4. Phonon dispersion curves, phonon density of states (PHDOS) projected on C, S, H in CH$_4$ and H in H$_3$S, and Eliashberg spectral function $\alpha^2F(\omega)$ together with the electron-phonon integral $\lambda$ for (a) $Cmc2_1$ at 150 GPa and (b) $P3m1$ at 50 GPa.

By comparing the electron-phonon integral λ of H$_3$S (see right panel of Fig. 5 in ref.[9]) and CS$_2$H$_{10}$ (see right panel of Fig. 4a and b), we can find that the contribution of S in the low frequency region (< 700 cm$^{-1}$) to $\lambda$ is basically the same. The decrease of $\lambda$ in $CS_2H_{10}$ is mainly due to the decrease in the contribution of H on the $H_3S$ in the middle frequency region. The phonons in the high frequency region (> 3000 cm$^{-1}$) mainly come from H on $CH_4$, which is consistent with $CSH_7$. Compared to $CSH_7$, $CS_2H_{10}$ has an additional peak contributed by H on $H_3S$ in 2000 - 2500 cm$^{-1}$ region. In *Cmc*2$_1$, this peak corresponds to the two H on the $H_3S$ closest to $CH_4$ (see H3 in Fig. 5a). The insertion of $CH_4$ breaks the S-H bond on one side, so they only connect one S atom, and their frequency is higher than the frequency of the H connecting two S atoms (H2). In *P*3*m*1, this peak corresponds to the three H on the $H_3S$ closest to $CH_4$ (see H3 in Fig. 5b). The insertion of $CH_4$ break the original S-H bond, and increase the frequency of the surrounding H (H3). Therefore, in these two types of $CS_2H_{10}$, the insertion of $CH_4$ affects the surrounding H on the $H_3S$ (H3), which weakens its effect on electron-phonon coupling, and only H protected by two S-H covalent bonds (H2) makes a major contribution to superconductivity. This is the reason why the $T_c$s of *Cmc*2$_1$-$CS_2H_{10}$ and *P*3*m*1-$CS_2H_{10}$ are lower than that of *Im*-3*m*-$H_3S$ and *R*3*m*-$H_3S$. The $CH_4$ in *R*3*m*-$CSH_7$ does not destroy the original S-H bond, and the surrounding H on the $H_3S$ (H2) is well protected by two S-H bonds[34]. So, the $T_c$ of *R*3*m*-$CSH_7$ can be as high as 181 - 191 K, which is very close to *Im*-3*m*-$H_3S$ (191 - 204 K). Therefore, in the future, when inserting $CH_4$ to reduce the pressure required for the stability of the $H_3S$ system, special attention should be paid to the position of $CH_4$ insertion, which will have a great impact on $T_c$.

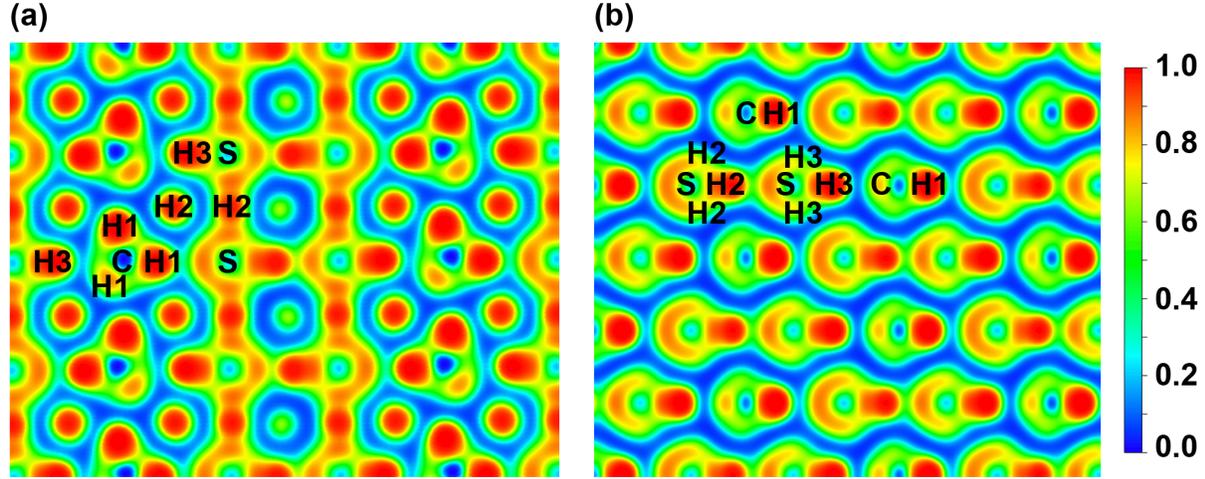

Fig. 5. Calculated electronic localization functions (ELF) of (a) $Cmc2_1$ at 150 GPa and (b) $P3m1$ at 50 GPa. H1 is the H on $CH_4$, and H2 is the H farther from $CH_4$ on $H_3S$. H3 is the H on $H_3S$ that is closer to $CH_4$.

To explore the bonding in the $Cmc2_1$ and $P3m1$ phases, their ELFs were plotted in Fig. 5. The ELF is useful for visualizing covalent bonds and lone pairs; it maps values in the range from 0 to 1, where 1 corresponds to perfect localization of the valence electrons indicative of a strong covalent bond. In $Cmc2_1$, the ELF value inside the $H_3S$ lattice is between 0.8-0.9, which has strong covalent properties like $Im$-$3m$-$H_3S$. However, the ELF value around $CH_4$ is close to 0, and it can only reach 0.4 between the nearest S atom, which is closer to the ELF value of the ionic bond. By comparing the ELF of $Cmc2_1$ (see Fig. 5a) and $Im$-$3m$-$H_3S$ (see Fig. 6d in Ref.[9]), we can find that the insertion of $CH_4$ does not affect H2 with two S-H bonds. But for H3, where one of the S-H bonds is broken, their charge density increases. By comparing the ELF of $P3m1$ (see Fig. 5b) and $R3m$-$H_3S$ (see Fig. 6b in Ref.[9]), we can also find that H3, which is closer to $CH_4$, has a more localized charge, while the charge distribution of H2 is similar to that of H in $R3m$-$H_3S$. Although this change caused by $CH_4$ is detrimental to superconductivity, it can make the $H_3S$ lattice stable at lower pressures. The volume of $CS_2H_{10}$ changes more with pressure than $H_3S$ (see Fig. 6). A smaller volume means a smaller PV term under high pressure which in turn means a lower enthalpy. The insertion of $CH_4$ breaks the isotropy of $H_3S$. The $H_3S$ lattice in $CS_2H_{10}$ still maintains the same properties as $H_3S$, but some H atoms were driven away from their symmetric positions to asymmetric ones in $H_3S$ lattice due to the anisotropic interaction between the $CH_4$ and $H_3S$ lattices. The lattice distortion caused by the movement of H atoms make $CS_2H_{10}$ more adaptable to pressure changes than $H_3S$. This is also the same in $I$-$43m$-$CSH_7$[33].

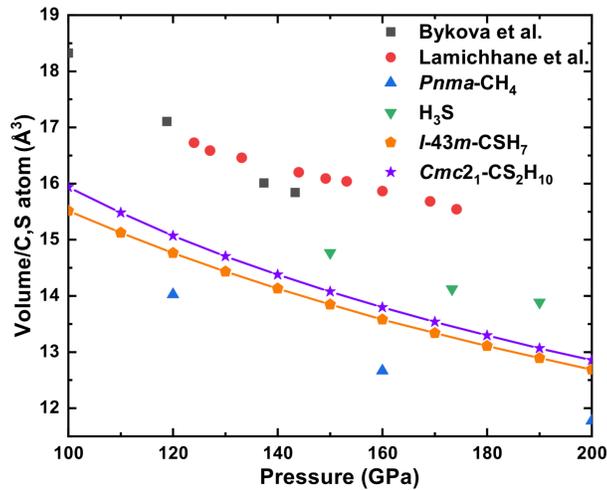

Fig. 6. Pressure-volume (P-V) relations measured for $CS_2H_{10}$ ($Cmc2_1$) compared to C-S-H compounds[31, 32], $H_3S$[12], $CH_4$[47] and $CSH_7$ ($I$-$43m$)[33, 34].

Due to the larger crystal lattice of $H_3S$, $CS_2H_{10}$ with more S has a larger volume than $CSH_7$. However, the volume of the experimentally synthesized C-S-H compound far exceeds that of $CSH_7$, $CS_2H_{10}$, or even $H_3S$, which means that the experimentally obtained structure is unlikely to be $CH_4$ insertion $H_3S$ or doped $H_3S$ structure. It is necessary to find a structure consistent with the experiment from more hydrogen-rich phases.

## Conclusions

We searched for the crystal structures of the $CS_2H_{10}$ by random structure searching up to 300 GPa. Four new and metastable phases $Cmc2_1$, $P3m1$, $P\text{-}3m1$ and $Pm$ were uncovered, which are all metallic and show superconducting properties. The $P3m1$ phase can be dynamically stable at much lower pressure of 50 GPa with $T_c$ of 155 K. For $Cmc2_1$ phase, it has high $T_c$ of 155 K at 150 GPa. The insertion of $CH_4$ greatly reduces the pressure required for the stability of the original $H_3S$ lattice, but it also has a negative impact on superconductivity that cannot be ignored. By comparing the experimental P-V diagram, we think that the phase obtained by the experiment is neither $CH_4$ insertion $H_3S$ nor doped $H_3S$. The C-S-H compound in the experiment should be found from a more hydrogen-rich area.

## Author Contributions

Defang Duan and Tian Cui directed the study. Mingyang Du conceived and performed the main work. All authors reviewed the paper.

## Conflict of interest

The authors declare no competing financial interest.

## ACKNOWLEDGMENT

This work was supported by the National Natural Science Foundation of China (Nos. 11674122, 51632002 and 52072188). Parts of calculations were performed in the High Performance Computing Center (HPCC) of Jilin University and TianHe-1(A) at the National Supercomputer Center in Tianjin.